\documentclass[journal]{IEEEtran}

\usepackage{epsfig,amsmath,amssymb,epsf,amsthm,scalefnt,multirow}
\usepackage{xcolor}
\usepackage{float}
\usepackage{cite}
\usepackage{psfrag}
\usepackage{algorithm}
\usepackage{algpseudocode}
\algrenewcommand\algorithmicrequire{\textbf{Input:}}
\algrenewcommand\algorithmicensure{\textbf{Output:}}

\newtheorem*{lemma*}{Lemma}
\newtheorem{observation}{Observation}

\newtheorem{conjecture}{Conjecture}

\begin{document}

\title{Channel Estimation via Gradient Pursuit for MmWave Massive MIMO Systems with One-Bit ADCs}

\author{In-soo Kim and Junil Choi
\thanks{The authors are with the School of Electrical Engineering, KAIST, Daejeon, Korea (e-mail: insookim@kaist.ac.kr; junil@kaist.ac.kr).}
\thanks{This work was partly supported by Institute for Information \& communications Technology Promotion(IITP) grant funded by the Korea government(MSIT) (No. 2016-0-00123, Development of Integer-Forcing MIMO Transceivers for 5G \& Beyond Mobile Communication Systems) and by the National Research Foundation (NRF) grant funded by the MSIT of the Korea government (2019R1C1C1003638).}}

\maketitle

\begin{abstract}
In millimeter wave (mmWave) massive multiple-input multiple-output (MIMO) systems, one-bit analog-to-digital converters (ADCs) are employed to reduce the impractically high power consumption, which is incurred by the wide bandwidth and large arrays. In practice, the mmWave band consists of a small number of paths, thereby rendering sparse virtual channels. Then, the resulting maximum a posteriori (MAP) channel estimation problem is a sparsity-constrained optimization problem, which is NP-hard to solve. In this paper, iterative approximate MAP channel estimators for mmWave massive MIMO systems with one-bit ADCs are proposed, which are based on the gradient support pursuit (GraSP) and gradient hard thresholding pursuit (GraHTP) algorithms. The GraSP and GraHTP algorithms iteratively pursue the gradient of the objective function to approximately optimize convex objective functions with sparsity constraints, which are the generalizations of the compressive sampling matching pursuit (CoSaMP) and hard thresholding pursuit (HTP) algorithms, respectively, in compressive sensing (CS). However, the performance of the GraSP and GraHTP algorithms is not guaranteed when the objective function is ill-conditioned, which may be incurred by the highly coherent sensing matrix. In this paper, the band maximum selecting (BMS) hard thresholding technique is proposed to modify the GraSP and GraHTP algorithms, namely the BMSGraSP and BMSGraHTP algorithms, respectively. The BMSGraSP and BMSGraHTP algorithms pursue the gradient of the objective function based on the band maximum criterion instead of the naive hard thresholding. In addition, a fast Fourier transform-based (FFT-based) fast implementation is developed to reduce the complexity. The BMSGraSP and BMSGraHTP algorithms are shown to be both accurate and efficient, whose performance is verified through extensive simulations.
\end{abstract}

\section{Introduction}
In millimeter wave (mmWave) massive multiple-input multiple-output (MIMO) systems, the wide bandwidth of the mmWave band in the range of 30-300 GHz offers a high data rate, which guarantees a significant performance gain \cite{6894453, 7306533, 7397887, 6736746}. However, the power consumption of analog-to-digital converters (ADCs) is scaled quadratically with the sampling rate and exponentially with the ADC resolution, which renders high-resolution ADCs impractical for mmWave systems \cite{1550190}. To reduce the power consumption, low-resolution ADCs were suggested as a possible solution, which recently gained popularity \cite{7600443, 7307134, 7420605, 7894211}. Coarsely quantizing the received signal using low-resolution ADCs results in an irreversible loss of information, which might cause a significant performance degradation. In this paper, we consider the extreme scenario of using one-bit ADCs for mmWave systems.

In practice, the mmWave band consists of a small number of propagation paths, which results in sparse virtual channels. In the channel estimation point of view, sparse channels are favorable because the required complexity and measurements can be reduced. Sparsity-constrained channel distributions, however, cannot be described in closed forms, which makes it difficult to exploit Bayesian channel estimation. In \cite{7439790, 7931630}, channel estimators for massive MIMO systems with one-bit ADCs were proposed, which account for the effect of the coarse quantization. The near maximum likelihood (nML) channel estimator \cite{7439790} selects the maximizer of the likelihood function subject to the $L^{2}$-norm constraint as the estimate of the channel, which is solved using the projected gradient descent method \cite{doi:10.1057/palgrave.jors.2600425}. However, the channel sparsity was not considered in \cite{7439790}. In \cite{7931630}, the Bussgang linear minimum mean squared error (BLMMSE) channel estimator was derived by linearizing one-bit ADCs based on the Bussgang decomposition \cite{bussgang1952crosscorrelation}. The BLMMSE channel estimator is an LMMSE channel estimator for massive MIMO systems with one-bit ADCs, whose assumption is that the channel is Gaussian. Therefore, the sparsity of the channel is not taken into account in \cite{7931630} either.

To take the channel sparsity into account, iterative approximate MMSE estimators for mmWave massive MIMO systems with one-bit ADCs were proposed in \cite{8320852, 8171203}. The generalized expectation consistent signal recovery (GEC-SR) algorithm in \cite{8320852} is an iterative approximate MMSE estimator based on the turbo principle \cite{7541826}, which can be applied to any nonlinear function of the linearly-mapped signal to be estimated. Furthermore, the only constraint on the distribution of the signal to be estimated is that its elements must be independent and identically distributed (i.i.d.) random variables. Therefore, the GEC-SR algorithm can be used as an approximate MMSE channel estimator for any ADC resolutions ranging from one-bit to high-resolution ADCs. However, the inverse of the sensing matrix is required at each iteration, which is impractical in massive MIMO systems in the complexity point of view.

The generalized approximate message passing-based (GAMP-based) channel estimator for mmWave massive MIMO systems with low-resolution ADCs was proposed in \cite{8171203}, which is another iterative approximate MMSE channel estimator. In contrast to the GEC-SR algorithm, only matrix-vector multiplications are required at each iteration, which is favorable in the complexity point of view. As in the GEC-SR algorithm, the GAMP-based algorithm can be applied to any ADC resolutions and any channel distributions as long as the elements of channel are i.i.d. random variable. The performance of the GEC-SR and GAMP algorithms, however, cannot be guaranteed when the sensing matrix is ill-conditioned, which frequently occurs in the mmWave band. To prevent the sensing matrix from becoming ill-conditioned, the GAMP-based channel estimator constructs the virtual channel representation using discrete Fourier transform (DFT) matrices, whose columns are orthogonal. However, such virtual channel representation results in a large gridding error, which leads to performance degradation.

Our goal is to propose an iterative approximate maximum a posteriori (MAP) channel estimator for mmWave massive MIMO systems with one-bit ADCs. Due to the sparse nature, the MAP channel estimation problem is converted to a sparsity-constrained optimization problem, which is NP-hard to solve \cite{JMLR:v14:bahmani13a}. To approximately solve such problems iteratively, the gradient support pursuit (GraSP) and gradient hard thresholding pursuit (GraHTP) algorithms were proposed in \cite{JMLR:v14:bahmani13a} and \cite{JMLR:v18:14-415}, respectively. The GraSP and GraHTP algorithms pursue the gradient of the objective function at each iteration by hard thresholding. These algorithms are the generalizations of the compressive sampling matching pursuit (CoSaMP) \cite{NEEDELL2009301} and hard thresholding pursuit (HTP) \cite{doi:10.1137/100806278} algorithms, respectively, in compressive sensing (CS).

With highly coherent sensing matrix, however, the GraSP and GraHTP algorithms do not perform appropriately since the objective function becomes ill-conditioned. To remedy such break down, we exploit the band maximum selecting (BMS) hard thresholding technique, which is then applied to the GraSP and GraHTP algorithms to propose the BMSGraSP and BMSGraHTP algorithms, respectively. The proposed BMS-based algorithms perform hard thresholding for the gradient of the objective function based on the proposed band maximum criterion, which tests whether an index is the ground truth index or the by-product of another index. To reduce the complexity of the BMS-based algorithms, a fast Fourier transform-based (FFT-based) fast implementation of the objective function and gradient is proposed. The BMS-based algorithms are shown to be both accurate and efficient, which is verified through extensive simulations.

The rest of this paper is organized as follows. In Section \ref{mmwave_massive_mimo_systems_with_one-bit_adcs}, mmWave massive MIMO systems with one-bit ADCs are described. In Section \ref{problem_formulation}, the MAP channel estimation framework is formulated. In Section \ref{channel_estimation_via_gradient_pursuit}, the BMS hard thresholding technique is proposed, which is applied to the GraSP and GraHTP algorithms. In addition, an FFT-based fast implementation is proposed. In Section \ref{simulation_results}, the results and discussion are presented, and the conclusions are followed in Section \ref{conclusion}.

\textbf{Notation:} $a$, $\mathbf{a}$, and $\mathbf{A}$ denote a scalar, vector, and matrix, respectively. $\|\mathbf{a}\|_{0}$, $\|\mathbf{a}\|_{1}$, and $\|\mathbf{a}\|$ represent the $L^{0}$-, $L^{1}$-, and $L^{2}$-norm of $\mathbf{a}$, respectively. $\|\mathbf{A}\|_{\mathrm{F}}$ is the Frobenius norm of $\mathbf{A}$. The transpose, conjugate transpose, and conjugate of $\mathbf{A}$ are denoted as $\mathbf{A}^{\mathrm{T}}$, $\mathbf{A}^{\mathrm{H}}$, and $\overline{\mathbf{A}}$, respectively. The element-wise vector multiplication and division of $\mathbf{a}$ and $\mathbf{b}$ are denoted as $\mathbf{a}\odot\mathbf{b}$ and $\mathbf{a}\oslash\mathbf{b}$, respectively. The sum of all of the elements of $\mathbf{a}$ is denoted as $\mathrm{sum}(\mathbf{a})$. The vectorization of $\mathbf{A}$ is denoted as $\mathrm{vec}(\mathbf{A})$, which is formed by stacking all of the columns of $\mathbf{A}$. The unvectorization of $\mathbf{a}$ is denoted as $\mathrm{unvec}(\mathbf{a})$, which is the inverse of $\mathrm{vec}(\mathbf{A})$. The Kronecker product of $\mathbf{A}$ and $\mathbf{B}$ is denoted as $\mathbf{A}\otimes\mathbf{B}$. The support of $\mathbf{a}$ is denoted as $\mathrm{supp}(\mathbf{a})$, which is the set of indices formed by collecting all of the indices of the nonzero elements of $\mathbf{a}$. The best $s$-term approximation of $\mathbf{a}$ is denoted as $\mathbf{a}|_{s}$, which is formed by leaving only the $s$ largest (in absolute value) elements of $\mathbf{a}$ and replacing the other elements with $0$. Similarly, the vector obtained by leaving only the elements of $\mathbf{a}$ indexed by the set $\mathcal{A}$ and replacing the other elements with $0$ is denoted as $\mathbf{a}|_{\mathcal{A}}$. The absolute value of a scalar $a$ and cardinality of a set $\mathcal{A}$ are denoted as $|a|$ and $|\mathcal{A}|$, respectively. The set difference between the sets $\mathcal{A}$ and $\mathcal{B}$, namely $\mathcal{A}\cap\mathcal{B}^{\mathrm{c}}$, is denoted as $\mathcal{A}\setminus\mathcal{B}$. $\phi(\mathbf{a})$ and $\Phi(\mathbf{a})$ are element-wise standard normal PDF and CDF functions of $\mathbf{a}$, whose $i$-th elements are $\frac{1}{\sqrt{2\pi}}e^{-\frac{a_{i}^{2}}{2}}$ and $\int_{-\infty}^{a_{i}}\frac{1}{\sqrt{2\pi}}e^{-\frac{x^{2}}{2}}dx$, respectively. The $m\times 1$ zero vector and $m\times m$ identity matrix are denoted as $\mathbf{0}_{m}$ and $\mathbf{I}_{m}$, respectively.

\section{mmWave Massive MIMO Systems with One-Bit ADCs}\label{mmwave_massive_mimo_systems_with_one-bit_adcs}
\subsection{System Model}
\begin{figure}[t]
\centering
\includegraphics[width=1\columnwidth]{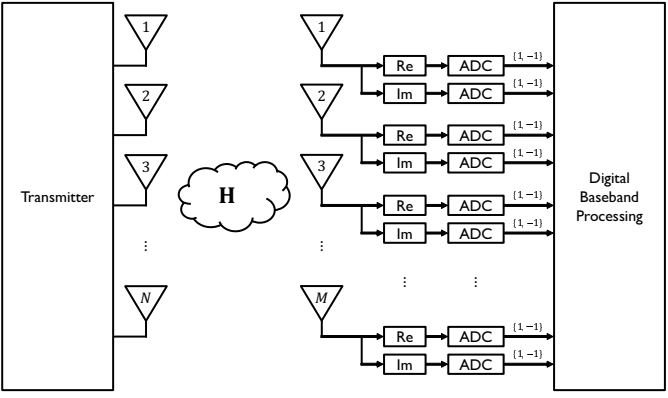}
\caption{A mmWave massive MIMO system with an $N$-antenna transmitter and $M$-antenna receiver. The real and imaginary parts of the received signal are quantized by one-bit ADCs.}\label{figure_1}
\end{figure}

As shown in Fig. \ref{figure_1}, consider a mmWave massive MIMO system with uniform linear arrays (ULAs) at the transmitter and receiver. The $N$-antenna transmitter transmits a training sequence of length $T$ to the $M$-antenna receiver. Therefore, the received signal $\mathbf{Y}=\begin{bmatrix}\mathbf{y}[1]&\mathbf{y}[2]&\cdots&\mathbf{y}[T]\end{bmatrix}\in\mathbb{C}^{M\times T}$ is
\begin{equation}\label{received_signal}
\mathbf{Y}=\sqrt{\rho}\mathbf{H}\mathbf{S}+\mathbf{N},
\end{equation}
which is the collection of the $t$-th received signal $\mathbf{y}[t]\in\mathbb{C}^{M}$ over $t\in\{1, \dots, T\}$. In the mmWave band, the channel $\mathbf{H}\in\mathbb{C}^{M\times N}$ consists of a small number of paths, whose parameters are the path gains, angle-of-arrivals (AoAs), and angle-of-departures (AoDs) \cite{6834753}. Therefore, $\mathbf{H}$ is
\begin{equation}\label{channel}
\mathbf{H}=\sum_{\ell=1}^{L}\alpha_{\ell}\mathbf{a}_{\mathrm{RX}}(\theta_{\mathrm{RX}, \ell})\mathbf{a}_{\mathrm{TX}}(\theta_{\mathrm{TX}, \ell})^{\mathrm{H}}
\end{equation}
where $L$ is the number of paths, $\alpha_{\ell}\in\mathbb{C}$ is the path gain of the $\ell$-th path, and $\theta_{\mathrm{RX}, \ell}\in[-\pi/2, \pi/2]$ and $\theta_{\mathrm{TX}, \ell}\in[-\pi/2, \pi/2]$ are the AoA and AoD of the $\ell$-th path, respectively. The steering vectors $\mathbf{a}_{\mathrm{RX}}(\theta_{\mathrm{RX}, \ell})\in\mathbb{C}^{M}$ and $\mathbf{a}_{\mathrm{TX}}(\theta_{\mathrm{TX}, \ell})\in\mathbb{C}^{N}$ are
\begin{align}
\mathbf{a}_{\mathrm{RX}}(\theta_{\mathrm{RX}, \ell})&=\frac{1}{\sqrt{M}}\begin{bmatrix}1&\cdots&e^{-j\pi(M-1)\sin(\theta_{\mathrm{RX}, \ell})}\end{bmatrix}^{\mathrm{T}},\\
\mathbf{a}_{\mathrm{TX}}(\theta_{\mathrm{TX}, \ell})&=\frac{1}{\sqrt{N}}\begin{bmatrix}1&\cdots&e^{-j\pi(N-1)\sin(\theta_{\mathrm{TX}, \ell})}\end{bmatrix}^{\mathrm{T}}
\end{align}
where the inter-element spacing is half-wavelength. The training sequence $\mathbf{S}=\begin{bmatrix}\mathbf{s}[1]&\mathbf{s}[2]&\cdots&\mathbf{s}[T]\end{bmatrix}\in\mathbb{C}^{N\times T}$ is the collection of the $t$-th training sequence $\mathbf{s}[t]\in\mathbb{C}^{N}$ over $t\in\{1, \dots, T\}$, whose power constraint is $\|\mathbf{s}[t]\|^{2}=N$. The additive white Gaussian noise (AWGN) $\mathbf{N}=\begin{bmatrix}\mathbf{n}[1]&\mathbf{n}[2]&\cdots&\mathbf{n}[T]\end{bmatrix}\in\mathbb{C}^{M\times T}$ is the collection of the $t$-th AWGN $\mathbf{n}[t]\in\mathbb{C}^{M}$ over $t\in\{1, \dots, T\}$, which is distributed as $\mathrm{vec}(\mathbf{N})\sim\mathcal{CN}(\mathbf{0}_{MT}, \mathbf{I}_{MT})$. The signal-to-noise ratio (SNR) is defined as $\rho$.

At the receiver, the real and imaginary parts of the received signal are quantized by one-bit ADCs. The quantized received signal $\hat{\mathbf{Y}}$ is
\begin{align}\label{quantized_received_signal}
\hat{\mathbf{Y}}&=\mathrm{Q}(\mathbf{Y})\notag\\
                &=\mathrm{Q}(\sqrt{\rho}\mathbf{H}\mathbf{S}+\mathbf{N})
\end{align}
where $\mathrm{Q}(\cdot)$ is the one-bit quantization function, whose threshold is zero. Therefore, $\mathrm{Q}(\mathbf{Y})$ is
\begin{equation}
\mathrm{Q}(\mathbf{Y})=\mathrm{sign}(\mathrm{Re}(\mathbf{Y}))+j\mathrm{sign}(\mathrm{Im}(\mathbf{Y}))
\end{equation}
where $\mathrm{sign}(\cdot)$ is the element-wise sign function. The goal is to estimate $\mathbf{H}$ by estimating $\{\alpha_{\ell}\}_{\ell=1}^{L}$, $\{\theta_{\mathrm{RX}, \ell}\}_{\ell=1}^{L}$, and $\{\theta_{\mathrm{TX}, \ell}\}_{\ell=1}^{L}$ from $\hat{\mathbf{Y}}$.

\subsection{Virtual Channel Representation}
In the mmWave channel model in \eqref{channel}, $\{\theta_{\mathrm{RX}, \ell}\}_{\ell=1}^{L}$ and $\{\theta_{\mathrm{TX}, \ell}\}_{\ell=1}^{L}$ are hidden in $\{\mathbf{a}_{\mathrm{RX}}(\theta_{\mathrm{RX}, \ell})\}_{\ell=1}^{L}$ and $\{\mathbf{a}_{\mathrm{TX}}(\theta_{\mathrm{TX}, \ell})\}_{\ell=1}^{L}$, respectively. The nonlinear mapping of $\{\theta_{\mathrm{RX}, \ell}\}_{\ell=1}^{L}$ and $\{\theta_{\mathrm{TX}, \ell}\}_{\ell=1}^{L}$ to $\mathbf{Y}$ renders a nonlinear channel estimation problem. To convert the nonlinear channel estimation problem to a linear channel estimation problem, we adopt the virtual channel representation \cite{1033686}.

The virtual channel representation of $\mathbf{H}$ is
\begin{equation}\label{virtual_channel_representation}
\mathbf{H}\approx\mathbf{A}_{\mathrm{RX}}\mathbf{X}^{*}\mathbf{A}_{\mathrm{TX}}^{\mathrm{H}}
\end{equation}
where the dictionary pair $\mathbf{A}_{\mathrm{RX}}\in\mathbb{C}^{M\times B_{\mathrm{RX}}}$ and $\mathbf{A}_{\mathrm{TX}}\in\mathbb{C}^{N\times B_{\mathrm{TX}}}$ is the collection of $B_{\mathrm{RX}}\geq M$ steering vectors and $B_{\mathrm{TX}}\geq N$ steering vectors, respectively. Therefore, $\mathbf{A}_{\mathrm{RX}}$ and $\mathbf{A}_{\mathrm{TX}}$ are
\begin{align}
\mathbf{A}_{\mathrm{RX}}&=\begin{bmatrix}\mathbf{a}_{\mathrm{RX}}(\omega_{\mathrm{RX}, 1})&\cdots&\mathbf{a}_{\mathrm{RX}}(\omega_{\mathrm{RX}, B_{\mathrm{RX}}})\end{bmatrix},\\
\mathbf{A}_{\mathrm{TX}}&=\begin{bmatrix}\mathbf{a}_{\mathrm{TX}}(\omega_{\mathrm{TX}, 1})&\cdots&\mathbf{a}_{\mathrm{TX}}(\omega_{\mathrm{TX}, B_{\mathrm{TX}}})\end{bmatrix},
\end{align}
whose gridding AoAs $\{\omega_{\mathrm{RX}, i}\}_{i=1}^{B_{\mathrm{RX}}}$ and AoDs $\{\omega_{\mathrm{TX}, j}\}_{j=1}^{B_{\mathrm{TX}}}$ are selected so as to form overcomplete DFT matrices. The gridding AoAs and AoDs are the $B_{\mathrm{RX}}$ and $B_{\mathrm{TX}}$ points from $[-\pi/2, \pi/2]$, respectively, to discretize the AoAs and AoDs because the ground truth AoAs and AoDs are unknown. To make a dictionary pair of the overcomplete DFT matrix form, the gridding AoAs and AoDs are given as $\omega_{\mathrm{RX}, i}=\sin^{-1}(-1+2/B_{\mathrm{RX}}\cdot (i-1))$ and $\omega_{\mathrm{RX}, j}=\sin^{-1}(-1+2/B_{\mathrm{TX}}\cdot (j-1))$, respectively. We prefer overcomplete DFT matrices because they are relatively well-conditioned, and DFT matrices are friendly to the FFT-based implementation, which will be discussed in Section \ref{channel_estimation_via_gradient_pursuit}. The virtual channel $\mathbf{X}^{*}\in\mathbb{C}^{B_{\mathrm{RX}}\times B_{\mathrm{TX}}}$ is the collection of $\{\alpha_{\ell}\}_{\ell=1}^{L}$, whose $(i, j)$-th element is $\alpha_{\ell}$ whenever $(\omega_{\mathrm{RX}, i}, \omega_{\mathrm{TX}, j})$ is the nearest to $(\theta_{\mathrm{RX}, \ell}, \theta_{\mathrm{TX}, \ell})$ but zero otherwise. In general, the error between $\mathbf{H}$ and $\mathbf{A}_{\mathrm{RX}}\mathbf{X}^{*}\mathbf{A}_{\mathrm{TX}}^{\mathrm{H}}$ is inversely proportional to $B_{\mathrm{RX}}$ and $B_{\mathrm{TX}}$. To approximate $\mathbf{H}$ using \eqref{virtual_channel_representation} with negligible error, the dictionary pair must be dense, namely $B_{\mathrm{RX}}\gg M$ and $B_{\mathrm{TX}}\gg N$.

Before we proceed, we provide a supplementary explanation on the approximation in \eqref{virtual_channel_representation}. In this work, we attempt to estimate the $L$-sparse $\mathbf{X}^{*}$ in \eqref{virtual_channel_representation} because the sparse assumption on $\mathbf{X}^{*}$ is favorable when the goal is to formulate the channel estimation problem as a sparsity-constrained problem. The cost of assuming that $\mathbf{X}^{*}$ is $L$-sparse is, as evident, the approximation error shown in \eqref{virtual_channel_representation}. Alternatively, the approximation error can be perfectly removed by considering $\mathbf{X}^{*}$ satisfying $\mathbf{H}=\mathbf{A}_{\mathrm{RX}}\mathbf{X}^{*}\mathbf{A}_{\mathrm{TX}}^{\mathrm{H}}$, i.e., equality instead of approximation. One well-known $\mathbf{X}^{*}$ satisfying the equality is the minimum Frobenius norm solution, i.e., $\mathbf{X}^{*}=\mathbf{A}_{\mathrm{RX}}^{\mathrm{H}}(\mathbf{A}_{\mathrm{RX}}\mathbf{A}_{\mathrm{RX}}^{\mathrm{H}})^{-1}\mathbf{H}(\mathbf{A}_{\mathrm{TX}}\mathbf{A}_{\mathrm{TX}}^{\mathrm{H}})^{-1}\mathbf{A}_{\mathrm{TX}}$. Such $\mathbf{X}^{*}$, however, has no evident structure to exploit in channel estimation, which is the reason why we assume that $\mathbf{X}^{*}$ is $L$-sparse at the cost of the approximation error in \eqref{virtual_channel_representation}.

In practice, the arrays at the transmitter and receiver are typically large to compensate the path loss in the mmWave band, whereas the number of line-of-sight (LOS) and near LOS paths is small \cite{6894454}. Therefore, $\mathbf{X}^{*}$ is sparse when the dictionary pair is dense because only $L$ elements among $B_{\mathrm{RX}}B_{\mathrm{TX}}$ elements are nonzero where $L\ll MN\ll B_{\mathrm{RX}}B_{\mathrm{TX}}$. In the sequel, we use the shorthand notation $B=B_{\mathrm{RX}}B_{\mathrm{TX}}$.

To facilitate the channel estimation framework, we vectorize \eqref{received_signal} and \eqref{quantized_received_signal} in conjunction with \eqref{virtual_channel_representation}. First, note that
\begin{equation}\label{mismatch}
\mathbf{Y}=\sqrt{\rho}\mathbf{A}_{\mathrm{RX}}\mathbf{X}^{*}\mathbf{A}_{\mathrm{TX}}^{\mathrm{H}}\mathbf{S}+\mathbf{N}+\mathbf{E}
\end{equation}
where the gridding error $\mathbf{E}\in\mathbb{C}^{M\times T}$ represents the mismatch in \eqref{virtual_channel_representation}.\footnote{In practice, $\mathbf{X}^{*}$ may be either approximately sparse or exactly sparse to formulate \eqref{mismatch}. If $\mathbf{X}^{*}$ is approximately sparse, the leakage effect is taken into account so the mismatch in \eqref{virtual_channel_representation} becomes zero, namely $\mathrm{vec}(\mathbf{E})=\mathbf{0}_{MT}$. In contrast, the mismatch in \eqref{virtual_channel_representation} must be taken into account with a nonzero $\mathbf{E}$ when $\mathbf{X}^{*}$ is exactly sparse. Fortunately, $\mathbf{E}$ is inversely proportional to $B_{\mathrm{RX}}$ and $B_{\mathrm{TX}}$. Therefore, we adopt the latter definition of $\mathbf{X}^{*}$ and propose our algorithm ignoring $\mathbf{E}$ assuming that $B_{\mathrm{RX}}\gg M$ and $B_{\mathrm{TX}}\gg N$. The performance degradation from $\mathbf{E}$ will become less as $B_{\mathrm{RX}}$ and $B_{\mathrm{TX}}$ become sufficiently large.} Then, the vectorized received signal $\mathbf{y}=\mathrm{vec}(\mathbf{Y})\in\mathbb{C}^{MT}$ is
\begin{equation}
\mathbf{y}=\sqrt{\rho}\mathbf{A}\mathbf{x}^{*}+\mathbf{n}+\mathbf{e}
\end{equation}
where
\begin{align}
    \mathbf{A}&=\mathbf{S}^{\mathrm{T}}\overline{\mathbf{A}}_{\mathrm{TX}}\otimes\mathbf{A}_{\mathrm{RX}}\notag\\
              &=\begin{bmatrix}\mathbf{a}_{1}&\mathbf{a}_{2}&\cdots&\mathbf{a}_{B}\end{bmatrix},\\
\mathbf{x}^{*}&=\mathrm{vec}(\mathbf{X}^{*})\notag\\
              &=\begin{bmatrix}x_{1}^{*}&x_{2}^{*}&\cdots&x_{B}^{*}\end{bmatrix}^{\mathrm{T}},\\
    \mathbf{n}&=\mathrm{vec}(\mathbf{N}),\\
    \mathbf{e}&=\mathrm{vec}(\mathbf{E}).
\end{align}
The vectorized quantized received signal $\hat{\mathbf{y}}=\mathrm{vec}(\hat{\mathbf{Y}})\in\mathbb{C}^{MT}$ is
\begin{align}\label{vectorized_quantized_received_signal}
\hat{\mathbf{y}}&=\mathrm{Q}(\mathbf{y})\notag\\
                &=\mathrm{Q}(\sqrt{\rho}\mathbf{A}\mathbf{x}^{*}+\mathbf{n}+\mathbf{e}).
\end{align}
The goal is to estimate $L$-sparse $\mathbf{x}^{*}$ from $\hat{\mathbf{y}}$.

\section{Problem Formulation}\label{problem_formulation}
In this section, we formulate the channel estimation problem using the MAP criterion. To facilitate the MAP channel estimation framework, the real counterparts of the complex forms in \eqref{vectorized_quantized_received_signal} are introduced. Then, the likelihood function of $\mathbf{x}^{*}$ is derived.

The real counterparts are the collections of the real and imaginary parts of the complex forms. Therefore, the real counterparts $\hat{\mathbf{y}}_{\mathrm{R}}\in\mathbb{R}^{2MT}$, $\mathbf{A}_{\mathrm{R}}\in\mathbb{R}^{2MT\times2B}$, and $\mathbf{x}_{\mathrm{R}}^{*}\in\mathbb{R}^{2B}$ are
\begin{align}
\hat{\mathbf{y}}_{\mathrm{R}}&=\begin{bmatrix}\mathrm{Re}(\hat{\mathbf{y}})^{\mathrm{T}}&\mathrm{Im}(\hat{\mathbf{y}})^{\mathrm{T}}\end{bmatrix}^{\mathrm{T}}\notag\\
                             &=\begin{bmatrix}\hat{y}_{\mathrm{R}, 1}&\hat{y}_{\mathrm{R}, 2}&\cdots&\hat{y}_{\mathrm{R}, 2MT}\end{bmatrix}^{\mathrm{T}},\\
      \mathbf{A}_{\mathrm{R}}&=\begin{bmatrix}\mathrm{Re}(\mathbf{A})&-\mathrm{Im}(\mathbf{A})\\\mathrm{Im}(\mathbf{A})&\mathrm{Re}(\mathbf{A})\end{bmatrix}\notag\\
                             &=\begin{bmatrix}\mathbf{a}_{\mathrm{R}, 1}&\mathbf{a}_{\mathrm{R}, 2}&\cdots&\mathbf{a}_{\mathrm{R}, 2MT}\end{bmatrix}^{\mathrm{T}},\\
  \mathbf{x}_{\mathrm{R}}^{*}&=\begin{bmatrix}\mathrm{Re}(\mathbf{x}^{*})^{\mathrm{T}}&\mathrm{Im}(\mathbf{x}^{*})^{\mathrm{T}}\end{bmatrix}^{\mathrm{T}}\notag\\
                             &=\begin{bmatrix}x_{\mathrm{R}, 1}^{*}&x_{\mathrm{R}, 2}^{*}&\cdots&x_{\mathrm{R}, 2B}^{*}\end{bmatrix}^{\mathrm{T}},
\end{align}
which are the collections of the real and imaginary parts of $\hat{\mathbf{y}}$, $\mathbf{A}$, and $\mathbf{x}^{*}$, respectively. In the sequel, we use the complex forms and the real counterparts interchangeably. For example, $\mathbf{x}^{*}$ and $\mathbf{x}_{\mathrm{R}}^{*}$ refer to the same entity. 

Before we formulate the likelihood function of $\mathbf{x}^{*}$, note that $\mathbf{e}$ is hard to analyze. However, $\mathbf{e}$ is negligible when the dictionary pair is dense. Therefore, we formulate the likelihood function of $\mathbf{x}^{*}$ without $\mathbf{e}$. The price of such oversimplification is negligible when $B_{\mathrm{RX}}\gg M$ and $B_{\mathrm{TX}}\gg N$, which is to be shown in Section \ref{simulation_results} where $\mathbf{e}\neq\mathbf{0}_{MT}$. To derive the likelihood function of $\mathbf{x}^{*}$, note that
\begin{equation}\label{distribution}
\sqrt{\rho}\mathbf{A}\mathbf{x}^{*}+\mathbf{n}\sim\mathcal{CN}(\sqrt{\rho}\mathbf{A}\mathbf{x}^{*}, \mathbf{I}_{MT})
\end{equation}
given $\mathbf{x}^{*}$. Then, from \eqref{distribution} in conjunction with \eqref{vectorized_quantized_received_signal}, the log-likelihood function $f(\mathbf{x})$ is \cite{7439790}
\begin{align}\label{log_likelihood_function}
f(\mathbf{x})&=\log\mathrm{Pr}\begin{bmatrix}\hat{\mathbf{y}}=\mathrm{Q}(\sqrt{\rho}\mathbf{A}\mathbf{x}+\mathbf{n})\mid\mathbf{x}\end{bmatrix}\notag\\
             &=\sum_{i=1}^{2MT}\log\Phi(\sqrt{2\rho}\hat{y}_{\mathrm{R}, i}\mathbf{a}_{\mathrm{R}, i}^{\mathrm{T}}\mathbf{x}_{\mathrm{R}}).
\end{align}

If the distribution of $\mathbf{x}^{*}$ is known, the MAP estimate of $\mathbf{x}^{*}$ is
\begin{equation}\label{true_map}
\underset{\mathbf{x}\in\mathbb{C}^{B}}{\mathrm{argmax}}\ (f(\mathbf{x})+g_{\mathrm{MAP}}(\mathbf{x}))
\end{equation}
where $g_{\mathrm{MAP}}(\mathbf{x})$ is the logarithm of the PDF of $\mathbf{x}^{*}$. In practice, however, $g_{\mathrm{MAP}}(\mathbf{x})$ is unknown. Therefore, we formulate the MAP channel estimation framework based on $\{\alpha_{\ell}\}_{\ell=1}^{L}$, $\{\theta_{\mathrm{RX}, \ell}\}_{\ell=1}^{L}$, and $\{\theta_{\mathrm{TX}, \ell}\}_{\ell=1}^{L}$ where we assume the followings:
\begin{enumerate}
\item $\alpha_{\ell}\sim\mathcal{CN}(0, 1)$ for all $\ell$
\item $\theta_{\mathrm{RX}, \ell}\sim\mathrm{unif}([-\pi/2, \pi/2])$ for all $\ell$
\item $\theta_{\mathrm{TX}, \ell}\sim\mathrm{unif}([-\pi/2, \pi/2])$ for all $\ell$
\item $\{\alpha_{\ell}\}_{\ell=1}^{L}$, $\{\theta_{\mathrm{RX}, \ell}\}_{\ell=1}^{L}$, and $\{\theta_{\mathrm{TX}, \ell}\}_{\ell=1}^{L}$ are independent.
\end{enumerate}
Then, the MAP estimate of $\mathbf{x}^{*}$ considering the channel sparsity is
\begin{equation}\label{map}
\underset{\mathbf{x}\in\mathbb{C}^{B}}{\mathrm{argmax}}\ (f(\mathbf{x})+g(\mathbf{x}))\enspace\text{s.t.}\enspace\|\mathbf{x}\|_{0}\leq L
\end{equation}
where $g(\mathbf{x})=-\|\mathbf{x}_{\mathrm{R}}\|^{2}$ is the logarithm of the PDF of $\mathcal{CN}(\mathbf{0}_{B}, \mathbf{I}_{B})$ ignoring the constant factor. However, note that only the optimization problems \eqref{true_map} and \eqref{map} are equivalent in the sense that their solutions are the same, not $g_{\mathrm{MAP}}(\mathbf{x})$ and $g(\mathbf{x})$. In the ML channel estimation framework, the ML estimate of $\mathbf{x}^{*}$ is
\begin{equation}\label{ml}
\underset{\mathbf{x}\in\mathbb{C}^{B}}{\mathrm{argmax}}\ f(\mathbf{x})\enspace\text{s.t.}\enspace\|\mathbf{x}\|_{0}\leq L.
\end{equation}
In the sequel, we focus on solving \eqref{map} because \eqref{map} reduces to \eqref{ml} when $g(\mathbf{x})=0$. In addition, we denote the objective function and the gradient in \eqref{map} as $h(\mathbf{x})$ and $\nabla h(\mathbf{x})$, respectively. Therefore,
\begin{align}
       h(\mathbf{x})&=f(\mathbf{x})+g(\mathbf{x}),\\
\nabla h(\mathbf{x})&=\nabla f(\mathbf{x})+\nabla g(\mathbf{x})\notag\\
                    &=\begin{bmatrix}\nabla h(x_{1})&\nabla h(x_{2})&\cdots&\nabla h(x_{B})\end{bmatrix}^{\mathrm{T}}
\end{align}
where the differentiation is with respect to $\mathbf{x}$.

\section{Channel Estimation via Gradient Pursuit}\label{channel_estimation_via_gradient_pursuit}
In this section, we propose the BMSGraSP and BMSGraHTP algorithms to solve \eqref{map}, which are the variants of the GraSP \cite{JMLR:v14:bahmani13a} and GraHTP \cite{JMLR:v18:14-415} algorithms, respectively. Then, an FFT-based fast implementation is proposed. In addition, we investigate the limit of the BMSGraSP and BMSGraHTP algorithms in the high SNR regime in one-bit ADCs.

\subsection{Proposed BMSGraSP and BMSGraHTP Algorithms}
Note that $h(\mathbf{x})$ in \eqref{map} is concave because $f(\mathbf{x})$ and $g(\mathbf{x})$ are the sums of the logarithms of $\Phi(\cdot)$ and $\phi(\cdot)$, respectively, which are log-concave \cite{boyd2004convex}. However, \eqref{map} is not a convex optimization problem because the sparsity constraint is not convex. Furthermore, solving \eqref{map} is NP-hard because of its combinatorial complexity. To approximately optimize convex objective functions with sparsity constraints iteratively by pursuing the gradient of the objective function, the GraSP and GraHTP algorithms were proposed in \cite{JMLR:v14:bahmani13a} and \cite{JMLR:v18:14-415}, respectively.

To solve \eqref{map}, the GraSP and GraHTP algorithms roughly proceed as follows at each iteration when $\mathbf{x}$ is the current estimate of $\mathbf{x}^{*}$ where the iteration index is omitted for simplicity. First, the best $L$-term approximation of $\nabla h(\mathbf{x})$ is computed, which is
\begin{equation}
T_{L}(\nabla h(\mathbf{x}))=\nabla h(\mathbf{x})|_{L}
\end{equation}
where $T_{L}(\cdot)$ is the $L$-term hard thresholding function. Here, $T_{L}(\cdot)$ leaves only the $L$ largest elements (in absolute value) of $\nabla h(\mathbf{x})$, and sets all the other remaining elements to $0$. Then, after the estimate of $\mathrm{supp}(\mathbf{x}^{*})$ is updated by selecting
\begin{equation}\label{support_estimation}
\mathcal{I}=\mathrm{supp}(T_{L}(\nabla h(\mathbf{x}))),
\end{equation}
i.e., $\mathcal{I}$ is the set of indices formed by collecting the $L$ indices of $\nabla h(\mathbf{x})$ corresponding to its $L$ largest elements (in absolute value), the estimate of $\mathbf{x}^{*}$ is updated by solving the following optimization problem
\begin{equation}
\underset{\mathbf{x}\in\mathbb{C}^{B}}{\mathrm{argmax}}\ h(\mathbf{x})\enspace\text{s.t.}\enspace\mathrm{supp}(\mathbf{x})\subseteq\mathcal{I},
\end{equation}
which can be solved using convex optimization because the support constraint is convex \cite{boyd2004convex}. The GraSP and GraHTP algorithms are the generalizations of the CoSaMP \cite{NEEDELL2009301} and HTP \cite{doi:10.1137/100806278} algorithms, respectively. This follows because the gradient of the squared error is the scaled proxy of the residual.

To solve \eqref{map} using the GraSP and GraHTP algorithms, $h(\mathbf{x})$ is required either to have a stable restricted Hessian \cite{JMLR:v14:bahmani13a} or to be strongly convex and smooth \cite{JMLR:v18:14-415}. These conditions are the generalizations of the restricted isometry property (RIP) in CS \cite{eldar2012compressed}, which means that $h(\mathbf{x})$ is likely to satisfy these conditions when $\mathbf{A}$ is either a restricted isometry, well-conditioned, or incoherent. In practice, however, $\mathbf{A}$ is highly coherent because the dictionary pair is typically dense to reduce the mismatch in \eqref{virtual_channel_representation}.

To illustrate how the GraSP and GraHTP algorithms fail to solve \eqref{map} when $\mathbf{A}$ is highly coherent, consider the real counterpart of $\nabla h(\mathbf{x})$. The real counterpart $\nabla h(\mathbf{x}_{\mathrm{R}})\in\mathbb{R}^{2B}$ is
\begin{align}\label{gradient}
 &\nabla h(\mathbf{x}_{\mathrm{R}})\notag\\
=&\begin{bmatrix}\mathrm{Re}(\nabla h(\mathbf{x}))^{\mathrm{T}}&\mathrm{Im}(\nabla h(\mathbf{x}))^{\mathrm{T}}\end{bmatrix}^{\mathrm{T}}\notag\\
=&\sum_{i=1}^{2MT}\lambda(\sqrt{2\rho}\hat{y}_{\mathrm{R}, i}\mathbf{a}_{\mathrm{R}, i}^{\mathrm{T}}\mathbf{x}_{\mathrm{R}})\sqrt{2\rho}\hat{y}_{\mathrm{R}, i}\mathbf{a}_{\mathrm{R}, i}-2\mathbf{x}_{\mathrm{R}},
\end{align}
which follows from $\nabla\log\Phi(\mathbf{a}_{\mathrm{R}}^{\mathrm{T}}\mathbf{x}_{\mathrm{R}})=\lambda(\mathbf{a}_{\mathrm{R}}^{\mathrm{T}}\mathbf{x}_{\mathrm{R}})\mathbf{a}_{\mathrm{R}}$ and $\nabla\|\mathbf{x}_{\mathrm{R}}\|^{2}=2\mathbf{x}_{\mathrm{R}}$ where $\lambda(\cdot)=\phi(\cdot)\oslash\Phi(\cdot)$ is the inverse Mills ratio function.\footnote{The element-wise vector division in the inverse Mills ratio function is meaningless because the arguments of the inverse Mills ratio function are scalars in \eqref{gradient}. The reason we use the element-wise vector division in the inverse Mills ratio function will become clear in \eqref{vectorized_gradient}, whose arguments are vectors.} Then, the following observation holds from directly computing $\nabla h(x_{i})$, whose real and imaginary parts are the $i$-th and $(i+B)$-th elements of $\nabla h(\mathbf{x}_{\mathrm{R}})$, respectively.
\begin{observation}\label{observation}
$\nabla h(x_{i})=\nabla h(x_{j})$ if $\mathbf{a}_{i}=\mathbf{a}_{j}$ and $x_{i}=x_{j}$.
\end{observation}

However, Observation \ref{observation} is meaningless because $\mathbf{a}_{i}\neq\mathbf{a}_{j}$ unless $i=j$. To establish a meaningful observation, consider the coherence between $\mathbf{a}_{i}$ and $\mathbf{a}_{j}$, which reflects the proximity between $\mathbf{a}_{i}$ and $\mathbf{a}_{j}$ according to \cite{doi:10.1137/110838509, 1715541}
\begin{equation}
\mu(i, j)=\frac{|\mathbf{a}_{i}^{\mathrm{H}}\mathbf{a}_{j}|}{\|\mathbf{a}_{i}\|\|\mathbf{a}_{j}\|}.
\end{equation}
Then, using the $\eta$-coherence band, which is \cite{doi:10.1137/110838509}
\begin{equation}
B_{\eta}(i)=\{j\mid\mu(i, j)\geq\eta\}
\end{equation}
where $\eta\in(0, 1)$, we establish the following conjecture when $\eta$ is sufficiently large.
\begin{conjecture}\label{conjecture}
$\nabla h(x_{i})\approx\nabla h(x_{j})$ if $j\in B_{\eta}(i)$ and $x_{i}=x_{j}$.
\end{conjecture}

\begin{figure}[t]
\centering
\includegraphics[width=1\columnwidth]{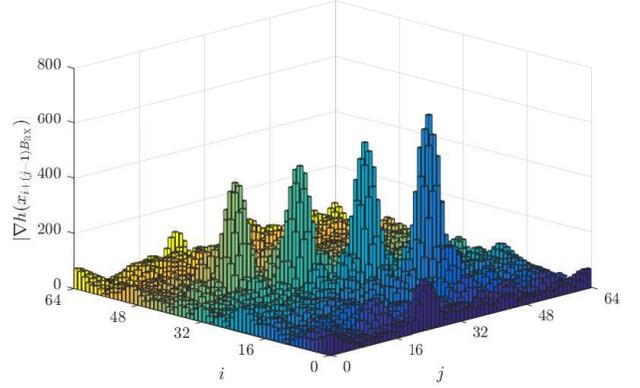}
\caption{The magnitude of $\mathrm{unvec}(\nabla h(\mathbf{x}))\in\mathbb{C}^{B_{\mathrm{RX}}\times B_{\mathrm{TX}}}$ at $\mathbf{x}=\mathbf{0}_{B}$, namely before hard thresholding.}\label{figure_2a}
\end{figure}

\begin{figure}[t]
\centering
\includegraphics[width=1\columnwidth]{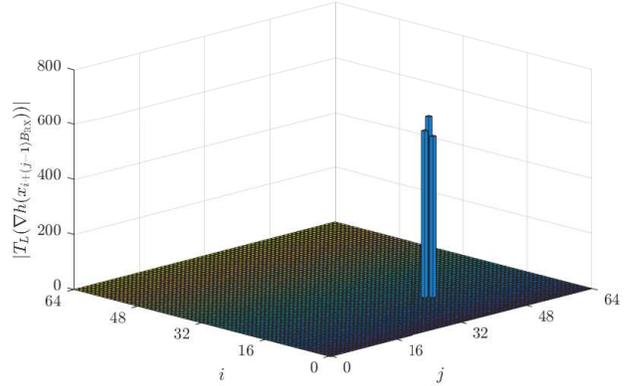}
\caption{The magnitude of $\mathrm{unvec}(T_{L}(\nabla h(\mathbf{x})))\in\mathbb{C}^{B_{\mathrm{RX}}\times B_{\mathrm{TX}}}$ at $\mathbf{x}=\mathbf{0}_{B}$, namely after hard thresholding. This shows how hard thresholding on $\nabla h(\mathbf{x})$ results in an incorrect estimate of $\mathrm{supp}(\mathbf{x}^{*})$ when $\mathbf{A}$ is highly coherent. In this example, $M=N=16$, $B_{\mathrm{RX}}=B_{\mathrm{TX}}=64$, $T=20$, $L=4$, $\mathrm{SNR}=20$ dB, and $\mathrm{supp}(\mathbf{x}^{*})$ is widely spread.}\label{figure_2b}
\end{figure}

At this point, we use Conjecture \ref{conjecture} to illustrate how the GraSP and GraHTP algorithms fail to estimate $\mathrm{supp}(\mathbf{x}^{*})$ from \eqref{support_estimation} by naive hard thresholding when $\mathbf{A}$ is highly coherent. To proceed, consider the following example, which assumes that $\mathbf{x}^{*}$ and $\hat{\mathbf{y}}$ are realized with $\mathbf{x}$ representing the current estimate of $\mathbf{x}^{*}$ so as to satisfy
\begin{enumerate}
\item $i=\underset{k\in\{1, 2, \dots, B\}}{\mathrm{argmax}}\ |x_{k}^{*}|$
\item $i=\underset{k\in\{1, 2, \dots, B\}}{\mathrm{argmax}}\ |\nabla h(x_{k})|$
\item $\mathcal{J}\cap\mathrm{supp}(\mathbf{x}^{*})=\emptyset$
\end{enumerate}
where $i$ is the index corresponding to the largest element of the ground truth\footnote{We use the term ``ground truth" to emphasize that the ground truth $\mathbf{x}^{*}$ is the true virtual channel which actually gives the quantized received signal $\hat{\mathbf{y}}$ from \eqref{vectorized_quantized_received_signal}, whereas $\mathbf{x}$ merely represents the point where $\nabla h(\mathbf{x})$ is computed to estimate $\mathrm{supp}(\mathbf{x}^{*})$ via hard thresholding.} virtual channel $\mathbf{x}^{*}$, and
\begin{equation}
\mathcal{J}=\{j\mid j\in B_{\eta}(i), x_{i}=x_{j}\}\setminus\{i\}
\end{equation}
is the by-product of $i$. Here, $\mathcal{J}$ is called the by-product of $i$ because
\begin{align}\label{conjecture_example}
|\nabla h(x_{j})|&\approx|\nabla h(x_{i})|\notag\\
                 &=\underset{k\in\{1, 2, \dots, B\}}{\mathrm{max}}\ |\nabla h(x_{k})|,
\end{align}
which follows from Conjecture \ref{conjecture}, holds despite $x_{j}^{*}=0$ for all $j\in\mathcal{J}$. In other words, the by-product of $i$ refers to the fact that $\nabla h(x_{i})$ and $\nabla h(x_{j})$ are indistinguishable for all $j\in\mathcal{J}$ according to \eqref{conjecture_example}, but the elements of $\mathbf{x}^{*}$ indexed by $\mathcal{J}$ are $0$ according to 3). Therefore, when we attempt to estimate $\mathrm{supp}(\mathbf{x}^{*})$ by hard thresholding $\nabla h(\mathbf{x})$, the indices in $\mathcal{J}$ will likely be erroneously selected as the estimate of $\mathrm{supp}(\mathbf{x}^{*})$ because $\nabla h(x_{j})$ and the maximum element of $\nabla h(\mathbf{x})$, which is $\nabla h(x_{i})$ according to 2), are indistinguishable for all $j\in\mathcal{J}$.

To illustrate how \eqref{support_estimation} cannot estimate $\mathrm{supp}(\mathbf{x}^{*})$ when $\mathbf{A}$ is highly coherent, consider another example where $\nabla h(\mathbf{x})$ and $T_{L}(\nabla h(\mathbf{x}))$ are shown in Figs. \ref{figure_2a} and \ref{figure_2b}, respectively. In this example, $\mathrm{supp}(\mathbf{x}^{*})$ is widely spread, whereas most of $\mathrm{supp}(T_{L}(\nabla h(\mathbf{x})))$ are in the coherence band of the index of the maximum element of $\nabla h(\mathbf{x})$. This shows that hard thresholding $\nabla h(\mathbf{x})$ is not sufficient to distinguish whether an index is the ground truth index or the by-product of another index. To solve this problem, we propose the BMS hard thresholding technique.

%

\begin{algorithm}[t]
\caption{BMS hard thresholding technique}\label{bms}
\begin{algorithmic}[1]
\Require $\mathbf{x}$, $\nabla h(\mathbf{x})$, $L$
\Ensure $T_{\mathrm{BMS}, L}(\nabla h(\mathbf{x}))$
\State $\mathcal{S}=\emptyset$, $\mathcal{I}=\{1, 2, \dots, B\}$
\While {$|\mathcal{S}|<L$}
\State $i=\underset{j\in\mathcal{I}}{\mathrm{argmax}}\ |\nabla h(x_{j})|$
\State $\mathcal{J}=\{j\mid j\in B_{\eta}(i), x_{i}=x_{j}\}\setminus\{i\}$
\If {$|\nabla h(x_{i})|>\underset{j\in\mathcal{J}}{\mathrm{max}}\ |\nabla h(x_{j})|$}
\State $\mathcal{S}=\mathcal{S}\cup\{i\}$
\EndIf
\State $\mathcal{I}=\mathcal{I}\setminus\{i\}$
\EndWhile
\State $T_{\mathrm{BMS}, L}(\nabla h(\mathbf{x}))=\nabla h(\mathbf{x})|_{\mathcal{S}}$
\end{algorithmic}
\end{algorithm}

The BMS hard thresholding function $T_{\mathrm{BMS}, L}(\cdot)$ is an $L$-term hard thresholding function, which is proposed based on Conjecture \ref{conjecture}. The BMS hard thresholding technique is presented in Algorithm \ref{bms}. Line 3 selects the index of the maximum element of $\nabla h(\mathbf{x})$ from the unchecked index set as the current index. Line 4 constructs the by-product testing set. Line 5 checks whether the current index is greater than the by-product testing set. In this paper, Line 5 is referred to as the band maximum criterion. If the band maximum criterion is satisfied, the current index is selected as the estimate of $\mathrm{supp}(\mathbf{x}^{*})$ in Line 6. Otherwise, the current index is not selected as the estimate of $\mathrm{supp}(\mathbf{x}^{*})$ because the current index is likely to be the by-product of another index rather than the ground truth index. Line 8 updates the unchecked index set.

Note that Algorithm \ref{bms} is a hard thresholding technique applied to $\nabla h(\mathbf{x})$. If the BMS hard thresholding technique is applied to $\mathbf{x}+\kappa\nabla h(\mathbf{x})$ where $\kappa$ is the step size, $\nabla h(\mathbf{x})$ is replaced by $\mathbf{x}+\kappa\nabla h(\mathbf{x})$ in the input, output, and Lines 3, 5, and 10 of Algorithm \ref{bms}. This can be derived using the same logic based on Conjecture \ref{conjecture}. Now, we propose the BMSGraSP and BMSGraHTP algorithms to solve \eqref{map}.

The BMSGraSP and BMSGraHTP algorithms are the variants of the GraSP and GraHTP algorithms, respectively. The difference between the BMS-based and non-BMS-based algorithms is that the hard thresholding function is $T_{\mathrm{BMS}, L}(\cdot)$ instead of $T_{L}(\cdot)$. The BMSGraSP and BMSGraHTP algorithms are presented in Algorithms \ref{bmsgrasp} and \ref{bmsgrahtp}, respectively. Lines 3, 4, and 5 of Algorithms \ref{bmsgrasp} and \ref{bmsgrahtp} roughly proceed based on the same logic. Line 3 computes the gradient of the objective function. Line 4 selects $\mathcal{I}$ from the support of the hard thresholded gradient of the objective function. Line 5 maximizes the objective function subject to the support constraint. This can be solved using convex optimization because the objective function and support constraint are concave and convex, respectively. In addition, $\mathbf{b}$ is hard thresholded in Line 6 of Algorithm \ref{bmsgrasp} because $\mathbf{b}$ is at most $3L$-sparse. A natural halting condition of Algorithms \ref{bmsgrasp} and \ref{bmsgrahtp} is to halt when the current and previous $\mathrm{supp}(\tilde{\mathbf{x}})$ are the same. The readers who are interested in a more in-depth analyses of the GraSP and GraHTP algorithms are referred to \cite{JMLR:v14:bahmani13a} and \cite{JMLR:v18:14-415}, respectively.

\begin{algorithm}[t]
\caption{BMSGraSP algorithm}\label{bmsgrasp}
\begin{algorithmic}[1]
\Require $h(\cdot)$, $L$
\Ensure $\tilde{\mathbf{x}}$
\State $\tilde{\mathbf{x}}=\mathbf{0}_{B}$
\While {halting condition}
\State $\mathbf{z}=\nabla h(\tilde{\mathbf{x}})$
\State $\mathcal{I}=\mathrm{supp}(T_{\mathrm{BMS}, 2L}(\mathbf{z}))\cup\mathrm{supp}(\tilde{\mathbf{x}})$
\State $\mathbf{b}=\underset{\mathbf{x}\in\mathbb{C}^{B}}{\mathrm{argmax}}\ h(\mathbf{x})\enspace\text{s.t.}\enspace\mathrm{supp}(\mathbf{x})\subseteq\mathcal{I}$
\State $\tilde{\mathbf{x}}=T_{L}(\mathbf{b})$
\EndWhile
\end{algorithmic}
\end{algorithm}

\begin{algorithm}[t]
\caption{BMSGraHTP algorithm}\label{bmsgrahtp}
\begin{algorithmic}[1]
\Require $h(\cdot)$, $L$
\Ensure $\tilde{\mathbf{x}}$
\State $\tilde{\mathbf{x}}=\mathbf{0}_{B}$
\While {halting condition}
\State $\mathbf{z}=\tilde{\mathbf{x}}+\kappa\nabla h(\tilde{\mathbf{x}})$
\State $\mathcal{I}=\mathrm{supp}(T_{\mathrm{BMS}, L}(\mathbf{z}))$
\State $\tilde{\mathbf{x}}=\underset{\mathbf{x}\in\mathbb{C}^{B}}{\mathrm{argmax}}\ h(\mathbf{x})\enspace\text{s.t.}\enspace\mathrm{supp}(\mathbf{x})\subseteq\mathcal{I}$
\EndWhile
\end{algorithmic}
\end{algorithm}

\textbf{Remark 1:} Instead of hard thresholding $\mathbf{b}$, we can solve
\begin{equation}\label{debiasing}
\tilde{\mathbf{x}}=\underset{\mathbf{x}\in\mathbb{C}^{B}}{\mathrm{argmax}}\ h(\mathbf{x})\enspace\text{s.t.}\enspace\mathrm{supp}(\mathbf{x})\subseteq\mathrm{supp}(T_{L}(\mathbf{b})),
\end{equation}
which is a convex optimization problem, to obtain $\tilde{\mathbf{x}}$ in Line 6 of Algorithm \ref{bmsgrasp}. This is the debiasing variant of Algorithm \ref{bmsgrasp} \cite{JMLR:v14:bahmani13a}. The advantage of the debiasing variant of Algorithm \ref{bmsgrasp} is a more accurate estimate of $\mathbf{x}^{*}$. However, the complexity is increased, which is incurred by solving \eqref{debiasing}.

\textbf{Remark 2:} In this paper, we assume that only $h(\mathbf{x})$ and $\nabla h(\mathbf{x})$ are required at each iteration to solve \eqref{map} using Algorithms \ref{bmsgrasp} and \ref{bmsgrahtp}, which can be accomplished when the first order method is used to solve convex optimization problems in Line 5 of Algorithms \ref{bmsgrasp} and \ref{bmsgrahtp}. An example of such first order method is the gradient descent method with the backtracking line search \cite{boyd2004convex}.

\subsection{Fast Implementation via FFT}
In practice, the complexity of Algorithms \ref{bmsgrasp} and \ref{bmsgrahtp} is demanding because $h(\mathbf{x})$ and $\nabla h(\mathbf{x})$ are required at each iteration, which are high-dimensional functions defined on $\mathbb{C}^{B}$ where $B\gg MN$. In recent works on channel estimation and data detection in the mmWave band \cite{8171203, 8320852, 7390019}, the FFT-based implementation is widely used because $\mathbf{H}$ can be approximated by \eqref{virtual_channel_representation} using overcomplete DFT matrices. In this paper, an FFT-based fast implementation of $h(\mathbf{x})$ and $\nabla h(\mathbf{x})$ is proposed, which is motivated by \cite{8171203, 8320852, 7390019}.

To facilitate the analysis, we convert the summations in $h(\mathbf{x})$ and $\nabla h(\mathbf{x}_{\mathrm{R}})$ to matrix-vector multiplications by algebraically manipulating \eqref{log_likelihood_function} and \eqref{gradient}. Then, we obtain
\begin{align}
 &h(\mathbf{x})\notag\\
=&\mathrm{sum}(\log\Phi(\sqrt{2\rho}\hat{\mathbf{y}}_{\mathrm{R}}\odot\mathbf{A}_{\mathrm{R}}\mathbf{x}_{\mathrm{R}}))-\|\mathbf{x}_{\mathrm{R}}\|^{2},\\
 &\nabla h(\mathbf{x}_{\mathrm{R}})\notag\\
=&\mathbf{A}_{\mathrm{R}}^{\mathrm{T}}(\lambda(\sqrt{2\rho}\hat{\mathbf{y}}_{\mathrm{R}}\odot\mathbf{A}_{\mathrm{R}}\mathbf{x}_{\mathrm{R}})\odot\sqrt{2\rho}\hat{\mathbf{y}}_{\mathrm{R}})-2\mathbf{x}_{\mathrm{R}}\label{vectorized_gradient}
\end{align}
where we see that the bottlenecks of $h(\mathbf{x})$ and $\nabla h(\mathbf{x})$ come from the matrix-vector multiplications involving $\mathbf{A}_{\mathrm{R}}$ and $\mathbf{A}_{\mathrm{R}}^{\mathrm{T}}$ resulting from the large size of $\mathbf{A}$. For example, the size of $\mathbf{A}$ is $5120\times 65536$ in Section \ref{simulation_results} where $M=N=64$, $B_{\mathrm{RX}}=B_{\mathrm{TX}}=256$, and $T=80$.

To develop an FFT-based fast implementation of the matrix-vector multiplications involving $\mathbf{A}_{\mathrm{R}}$ and $\mathbf{A}_{\mathrm{R}}^{\mathrm{T}}$, define $\mathbf{c}_{\mathrm{R}}\in\mathbb{R}^{2MT}$ as $\mathbf{c}_{\mathrm{R}}=\lambda(\sqrt{2\rho}\hat{\mathbf{y}}_{\mathrm{R}}\odot\mathbf{A}_{\mathrm{R}}\mathbf{x}_{\mathrm{R}})\odot\sqrt{2\rho}\hat{\mathbf{y}}_{\mathrm{R}}$ from \eqref{vectorized_gradient} with $\mathbf{c}\in\mathbb{C}^{MT}$ being the complex form of $\mathbf{c}_{\mathrm{R}}$. From the fact that
\begin{align}
             \mathbf{A}_{\mathrm{R}}\mathbf{x}_{\mathrm{R}}&=\begin{bmatrix}\mathrm{Re}(\mathbf{A}\mathbf{x})^{\mathrm{T}}&\mathrm{Im}(\mathbf{A}\mathbf{x})^{\mathrm{T}}\end{bmatrix}^{\mathrm{T}},\\
\mathbf{A}_{\mathrm{R}}^{\mathrm{T}}\mathbf{c}_{\mathrm{R}}&=\begin{bmatrix}\mathrm{Re}(\mathbf{A}^{\mathrm{H}}\mathbf{c})^{\mathrm{T}}&\mathrm{Im}(\mathbf{A}^{\mathrm{H}}\mathbf{c})^{\mathrm{T}}\end{bmatrix}^{\mathrm{T}},
\end{align}
we now attempt to compute $\mathbf{A}\mathbf{x}$ and $\mathbf{A}^{\mathrm{H}}\mathbf{c}$ via the FFT. Then, $\mathbf{A}\mathbf{x}$ and $\mathbf{A}^{\mathrm{H}}\mathbf{c}$ are unvectorized according to
\begin{align}
             \mathrm{unvec}(\mathbf{A}\mathbf{x})&=\mathbf{A}_{\mathrm{RX}}\mathbf{X}\mathbf{A}_{\mathrm{TX}}^{\mathrm{H}}\mathbf{S}\notag\\
                                                 &=\underbrace{\mathbf{A}_{\mathrm{RX}}(\underbrace{\mathbf{S}^{\mathrm{H}}(\underbrace{\mathbf{A}_{\mathrm{TX}}\mathbf{X}^{\mathrm{H}}}_{\mathrm{FFT}})}_{\mathrm{IFFT}})^{\mathrm{H}}}_{\mathrm{FFT}},\label{first_matrix_multiplication}\\
\mathrm{unvec}(\mathbf{A}^{\mathrm{H}}\mathbf{c})&=\mathbf{A}_{\mathrm{RX}}^{\mathrm{H}}\mathbf{C}\mathbf{S}^{\mathrm{H}}\mathbf{A}_{\mathrm{TX}}\notag\\
                                                 &=\underbrace{\mathbf{A}_{\mathrm{RX}}^{\mathrm{H}}(\underbrace{\mathbf{A}_{\mathrm{TX}}^{\mathrm{H}}(\underbrace{\mathbf{S}\mathbf{C}^{\mathrm{H}}}_{\mathrm{FFT}})}_{\mathrm{IFFT}})^{\mathrm{H}}}_{\mathrm{IFFT}}\label{second_matrix_multiplication}
\end{align}
where $\mathbf{X}=\mathrm{unvec}(\mathbf{x})\in\mathbb{C}^{B_{\mathrm{RX}}\times B_{\mathrm{TX}}}$ and $\mathbf{C}=\mathrm{unvec}(\mathbf{c})\in\mathbb{C}^{M\times T}$. If the matrix multiplication involving $\mathbf{S}$ can be implemented using the FFT, e.g., Zadoff-Chu (ZC) \cite{1054840} or DFT \cite{7931630} training sequence, \eqref{first_matrix_multiplication} and \eqref{second_matrix_multiplication} can be implemented using the FFT because $\mathbf{A}_{\mathrm{RX}}$ and $\mathbf{A}_{\mathrm{TX}}$ are overcomplete DFT matrices. For example, each column of $\mathbf{A}_{\mathrm{TX}}\mathbf{X}^{\mathrm{H}}$ in \eqref{first_matrix_multiplication} can be computed using the $B_{\mathrm{TX}}$-point FFT with pruned outputs, i.e., retaining only $N$ outputs, because we constructed $\mathbf{A}_{\mathrm{TX}}$ as an overcomplete DFT matrix.

In particular, the matrix multiplications involving $\mathbf{A_{\mathrm{TX}}}$, $\mathbf{S}^{\mathrm{H}}$, and $\mathbf{A}_{\mathrm{RX}}$ in \eqref{first_matrix_multiplication} can be implemented with $B_{\mathrm{TX}}$-point FFT with pruned outputs repeated $B_{\mathrm{RX}}$ times, $T$-point IFFT with pruned inputs repeated $B_{\mathrm{RX}}$ times, and $B_{\mathrm{RX}}$-point FFT with pruned outputs repeated $T$ times, respectively.\footnote{The inputs and outputs are pruned because $\mathbf{A}_{\mathrm{RX}}$, $\mathbf{A}_{\mathrm{TX}}$, and $\mathbf{S}$ are rectangular, not square. The details of the pruned FFT are presented in \cite{1162205, 1162782, 1163246}.} Using the same logic, the matrix multiplications involving $\mathbf{S}$, $\mathbf{A}_{\mathrm{TX}}^{\mathrm{H}}$, and $\mathbf{A}_{\mathrm{RX}}^{\mathrm{H}}$ in \eqref{second_matrix_multiplication} can be implemented using $T$-point FFT with pruned outputs repeated $M$ times, $B_{\mathrm{TX}}$-point IFFT with pruned inputs repeated $M$ times, and $B_{\mathrm{RX}}$-point IFFT with pruned inputs repeated $B_{\mathrm{TX}}$ times, respectively. Therefore, the complexity of the FFT-based implementation of \eqref{first_matrix_multiplication} and \eqref{second_matrix_multiplication} is $O(B_{\mathrm{RX}}B_{\mathrm{TX}}\log B_{\mathrm{TX}}+B_{\mathrm{RX}}T\log T+TB_{\mathrm{RX}}\log B_{\mathrm{RX}})$ and $O(MT\log T+MB_{\mathrm{TX}}\log B_{\mathrm{TX}}+B_{\mathrm{TX}}B_{\mathrm{RX}}\log B_{\mathrm{RX}})$, respectively.

To illustrate the efficiency of the FFT-based implementation of \eqref{first_matrix_multiplication} and \eqref{second_matrix_multiplication}, $M/N$, $M/B_{\mathrm{RX}}$, $M/B_{\mathrm{TX}}$, and $M/T$ are assumed to be fixed. Then, the complexity of the FFT-based implementation of $\mathbf{A}\mathbf{x}$ and $\mathbf{A}^{\mathrm{H}}\mathbf{c}$ is $O(M^{2}\log M)$, whereas the complexity of directly computing $\mathbf{A}\mathbf{x}$ and $\mathbf{A}^{\mathrm{H}}\mathbf{c}$ is $O(M^{4})$. Therefore, the complexity of Algorithms \ref{bmsgrasp} and \ref{bmsgrahtp} is reduced when $h(\mathbf{x})$ and $\nabla h(\mathbf{x})$ are implemented using the FFT operations.

\textbf{Remark 3:} Line 5 of Algorithms \ref{bmsgrasp} and \ref{bmsgrahtp} is equivalent to solving
\begin{equation}\label{line_5}
\underset{\mathbf{x}_{\mathcal{I}}\in\mathbb{C}^{|\mathcal{I}|}}{\mathrm{argmax}}\ h_{\mathcal{I}}(\mathbf{x}_{\mathcal{I}})=\underset{\mathbf{x}_{\mathcal{I}}\in\mathbb{C}^{|\mathcal{I}|}}{\mathrm{argmax}}\ (f_{\mathcal{I}}(\mathbf{x}_{\mathcal{I}})+g_{\mathcal{I}}(\mathbf{x}_{\mathcal{I}}))
\end{equation}
where
\begin{align}
f_{\mathcal{I}}(\mathbf{x}_{\mathcal{I}})&=\log\mathrm{Pr}\begin{bmatrix}\hat{\mathbf{y}}=\mathrm{Q}(\sqrt{\rho}\mathbf{A}_{\mathcal{I}}\mathbf{x}_{\mathcal{I}}+\mathbf{n})\mid\mathbf{x}_{\mathcal{I}}\end{bmatrix},\\
g_{\mathcal{I}}(\mathbf{x}_{\mathcal{I}})&=-\|\mathbf{x}_{\mathcal{I}}\|^{2},
\end{align}
and $\mathbf{A}_{\mathcal{I}}\in\mathbb{C}^{MT\times |\mathcal{I}|}$ is the collection of $\mathbf{a}_{i}$ with $i\in\mathcal{I}$. Therefore, only $h_{\mathcal{I}}(\mathbf{x}_{\mathcal{I}})$ and $\nabla h_{\mathcal{I}}(\mathbf{x}_{\mathcal{I}})$ are required in Line 5 of Algorithms \ref{bmsgrasp} and \ref{bmsgrahtp}, which are low-dimensional functions defined on $\mathbb{C}^{|\mathcal{I}|}$ where $|\mathcal{I}|=O(L)$. If $h_{\mathcal{I}}(\mathbf{x}_{\mathcal{I}})$ and $\nabla h_{\mathcal{I}}(\mathbf{x}_{\mathcal{I}})$ are computed based on the same logic in \eqref{first_matrix_multiplication} and \eqref{second_matrix_multiplication} but $\mathbf{A}$ replaced by $\mathbf{A}_{\mathcal{I}}$, the complexity of Algorithms \ref{bmsgrasp} and \ref{bmsgrahtp} is reduced further because the size of the FFT is reduced in Line 5.

\section{Results and Discussion}\label{simulation_results}
In this section, we evaluate the performance of Algorithms \ref{bmsgrasp} and \ref{bmsgrahtp} from different aspects in terms of the accuracy, achievable rate, and complexity. Throughout this section, we consider a mmWave massive MIMO system with one-bit ADCs, whose parameters are $M=N=64$ and $T=80$. The rest vary from simulation to simulation, which consist of $B_{\mathrm{RX}}$, $B_{\mathrm{TX}}$, and $L$. In addition, we consider $\mathbf{S}$, whose rows are the circular shifts of the ZC training sequence of length $T$ as in \cite{8171203, 8310593}. Furthermore, $\mathbf{H}$ is either random or deterministic. If $\mathbf{H}$ is random, $\alpha_{\ell}\sim\mathcal{CN}(0, 1)$, $\theta_{\mathrm{RX}, \ell}\sim\mathrm{unif}([-\pi/2, \pi/2])$, and $\theta_{\mathrm{TX}, \ell}\sim\mathrm{unif}([-\pi/2, \pi/2])$ are independent. Otherwise, we consider different $\mathbf{H}$ from simulation to simulation.

The MSEs of $\{\alpha_{\ell}\}_{\ell=1}^{L}$, $\{\theta_{\mathrm{RX}, \ell}\}_{\ell=1}^{L}$, and $\{\theta_{\mathrm{TX}, \ell}\}_{\ell=1}^{L}$ are
\begin{align}
             \mathrm{MSE}(\{\alpha_{\ell}\}_{\ell=1}^{L})&=\mathbb{E}\left\{\frac{1}{L}\sum_{\ell=1}^{L}|\tilde{\alpha}_{\ell}-\alpha_{\ell}|^{2}\right\},\label{MSE_alpha}\\
\mathrm{MSE}(\{\theta_{\mathrm{RX}, \ell}\}_{\ell=1}^{L})&=\mathbb{E}\left\{\frac{1}{L}\sum_{\ell=1}^{L}(\tilde{\theta}_{\mathrm{RX}, \ell}-\theta_{\mathrm{RX}, \ell})^{2}\right\},\label{MSE_theta_RX}\\
\mathrm{MSE}(\{\theta_{\mathrm{TX}, \ell}\}_{\ell=1}^{L})&=\mathbb{E}\left\{\frac{1}{L}\sum_{\ell=1}^{L}(\tilde{\theta}_{\mathrm{TX}, \ell}-\theta_{\mathrm{TX}, \ell})^{2}\right\}\label{MSE_theta_TX}
\end{align}
where $(\tilde{\alpha}_{\ell}, \tilde{\theta}_{\mathrm{RX}, \ell}, \tilde{\theta}_{\mathrm{TX}, \ell})$ corresponds to some nonzero element of $\tilde{\mathbf{X}}=\mathrm{unvec}(\tilde{\mathbf{x}})\in\mathbb{C}^{B_{\mathrm{RX}}\times B_{\mathrm{TX}}}$. The normalized MSE (NMSE) of $\mathbf{H}$ is
\begin{equation}\label{MSE_H}
\mathrm{NMSE}(\mathbf{H})=\mathbb{E}\left\{\frac{\|\tilde{\mathbf{H}}-\mathbf{H}\|_{\mathrm{F}}^{2}}{\|\mathbf{H}\|_{\mathrm{F}}^{2}}\right\}
\end{equation}
where $\tilde{\mathbf{H}}=\mathbf{A}_{\mathrm{RX}}\tilde{\mathbf{X}}\mathbf{A}_{\mathrm{TX}}$. In \eqref{MSE_alpha}-\eqref{MSE_H}, the symbol $\tilde{\hphantom{\mathbf{y}}}$ is used to emphasize that the quantity is an estimate.

\begin{figure}[t]
\centering
\includegraphics[width=1\columnwidth]{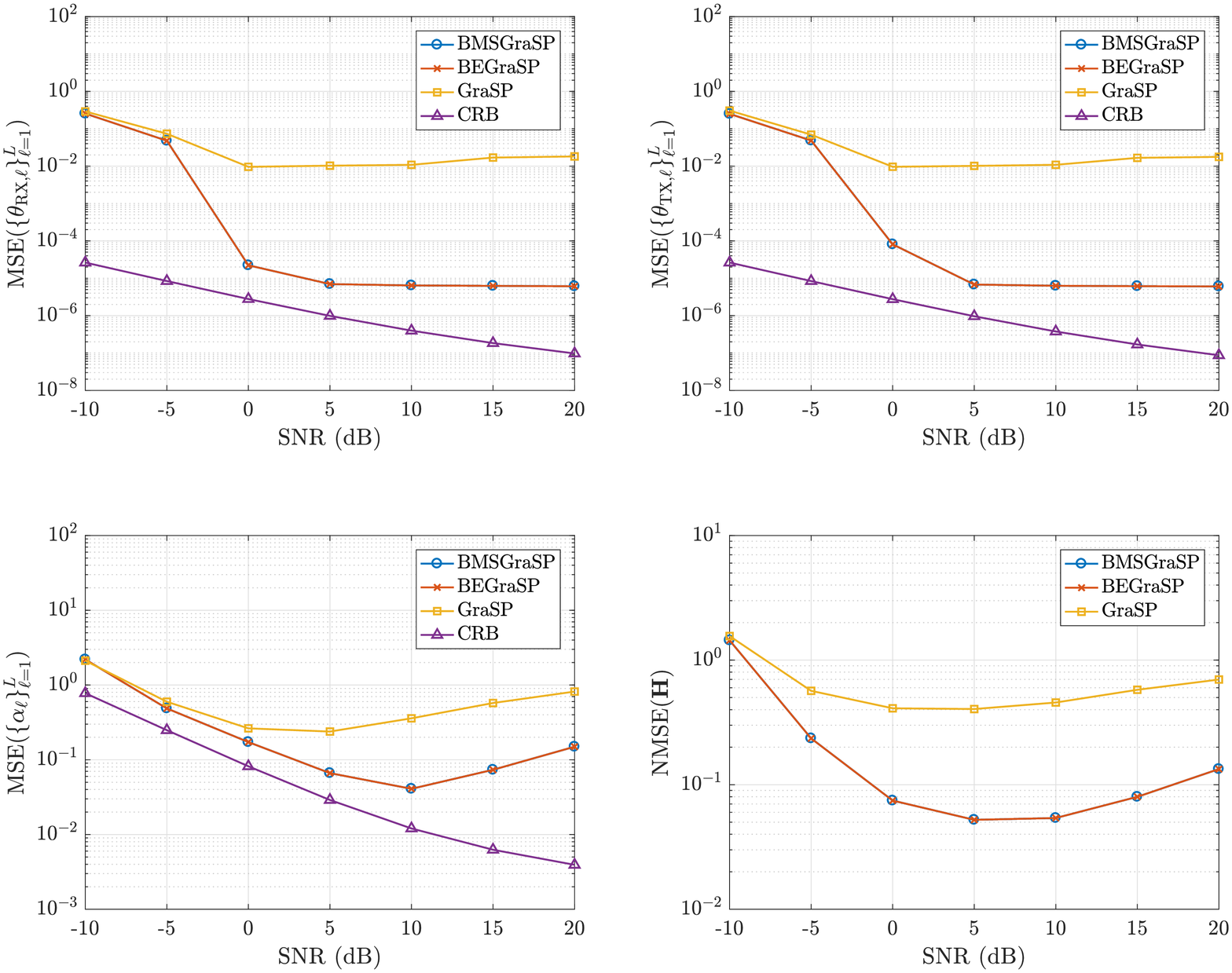}
\caption{MSEs of the BMS-based and BE-based GraSP algorithms for widely spread $\theta_{\mathrm{RX}, \ell}=\theta_{\mathrm{TX}, \ell}=\frac{\pi}{18}(\ell-1)$ with $M=N=64$, $T=80$, $L=8$, and $\alpha_{\ell}=(0.8+0.1(\ell-1))e^{j\frac{\pi}{4}(\ell-1)}$. The CRB is provided as a reference, which was derived in \cite{8335511}.}\label{figure_3a_alt}
\end{figure}

\begin{figure}[t]
\centering
\includegraphics[width=1\columnwidth]{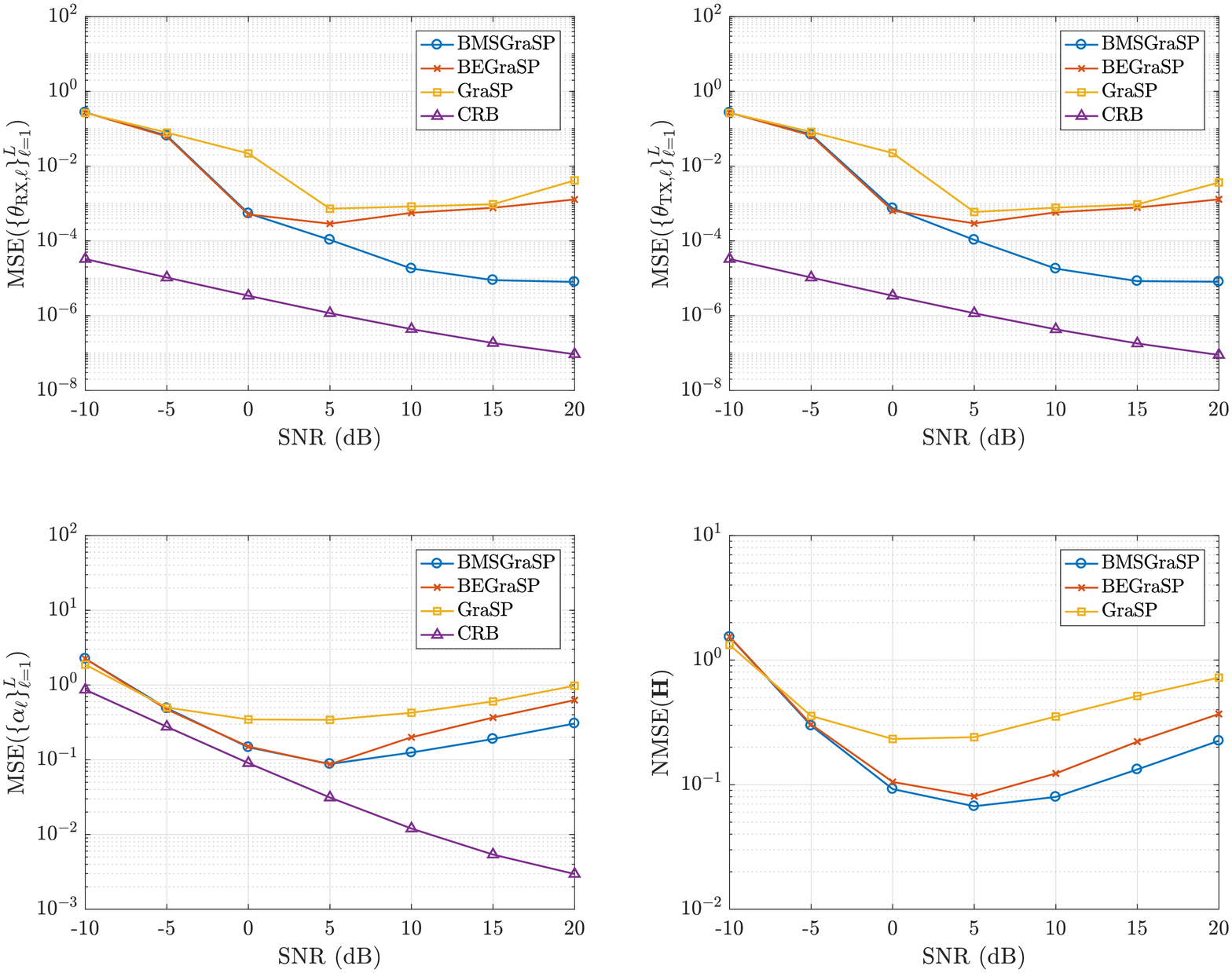}
\caption{MSEs of the BMS-based and BE-based GraSP algorithms for closely spread $\theta_{\mathrm{RX}, \ell}=\theta_{\mathrm{TX}, \ell}=\frac{\pi}{36}(\ell-1)$ with $M=N=64$, $T=80$, $L=8$, and $\alpha_{\ell}=(0.8+0.1(\ell-1))e^{j\frac{\pi}{4}(\ell-1)}$. The CRB is provided as a reference, which was derived in \cite{8335511}.}\label{figure_3b_alt}
\end{figure}

\begin{figure}[t]
\centering
\includegraphics[width=1\columnwidth]{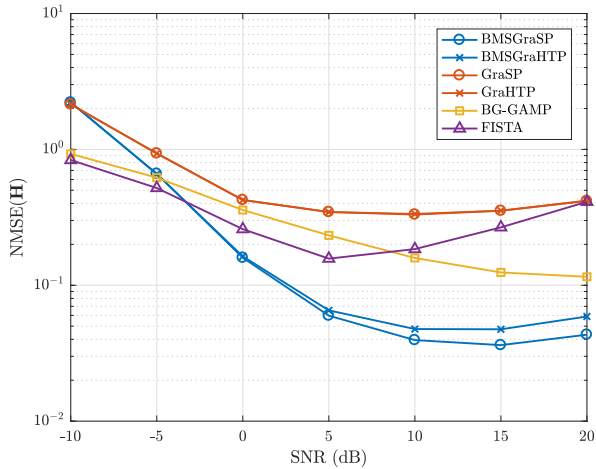}
\caption{NMSE vs. SNR where $M=N=64$, $T=80$, and $L=4$ with varying $B_{\mathrm{RX}}$ and $B_{\mathrm{TX}}$ from algorithm to algorithm.}\label{figure_4}
\end{figure}

\begin{figure}[t]
\centering
\includegraphics[width=1\columnwidth]{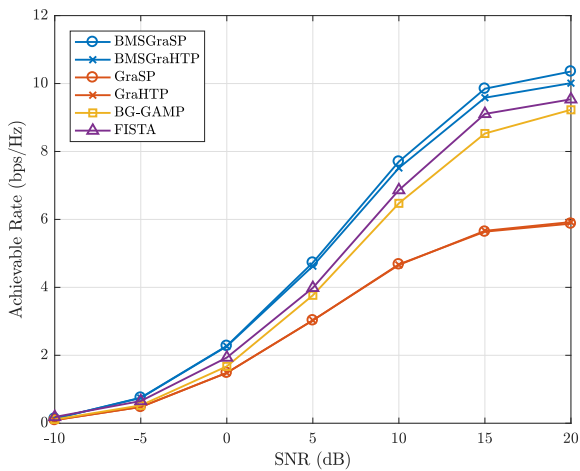}
\caption{Achievable rate lower bound \cite{8171203} vs. SNR where $M=N=64$, $T=80$, and $L=4$ with varying $B_{\mathrm{RX}}$ and $B_{\mathrm{TX}}$ from algorithm to algorithm.}\label{figure_5}
\end{figure}

Throughout this section, we consider the debiasing variant of Algorithm \ref{bmsgrasp}. The halting condition of Algorithms \ref{bmsgrasp} and \ref{bmsgrahtp} is to halt when the current and previous $\mathrm{supp}(\tilde{\mathbf{x}})$ are the same. The gradient descent method is used to solve convex optimization problems, which consist of \eqref{debiasing} and Line 5 of Algorithms \ref{bmsgrasp} and \ref{bmsgrahtp}. The backtracking line search is used to compute the step size in the gradient descent method and $\kappa$ in Line 3 of Algorithm \ref{bmsgrahtp}. In addition, $\eta$ is selected so that Conjecture \ref{conjecture} is satisfied. In this paper, we select the maximum $\eta$ satisfying
\begin{equation}\label{criterion}
\underset{i\in\{1, 2, \dots, B\}}{\mathrm{min}}\ |B_{\eta}(i)|>1.
\end{equation}
For example, the maximum $\eta$ satisfying \eqref{criterion} is $\eta=0.6367$ when $B_{\mathrm{RX}}=2M$ and $B_{\mathrm{TX}}=2N$. The channel estimation criterion of Algorithms \ref{bmsgrasp} and \ref{bmsgrahtp} is either MAP or ML, which depends on whether $\mathbf{H}$ is random or deterministic.
To compare the BMS-based and non-BMS-based algorithms, the performance of the GraSP and GraHTP algorithms is shown as a reference in Figs. \ref{figure_3a_alt}, \ref{figure_3b_alt}, \ref{figure_4}, and \ref{figure_5}. The GraSP and GraHTP algorithms forbid $B_{\mathrm{RX}}\gg M$ and $B_{\mathrm{TX}}\gg N$ because the GraSP and GraHTP algorithms diverge when $\mathbf{A}$ is highly coherent. Therefore, the parameters are selected as $B_{\mathrm{RX}}=M$ and $B_{\mathrm{TX}}=N$ when the GraSP and GraHTP algorithms are implemented. Such $B_{\mathrm{RX}}$ and $B_{\mathrm{TX}}$, however, are problematic because the mismatch in \eqref{virtual_channel_representation} is inversely proportional to $B_{\mathrm{RX}}$ and $B_{\mathrm{TX}}$.

In Figs. \ref{figure_3a_alt} and \ref{figure_3b_alt}, we compare the accuracy of the BMS-based and band excluding-based (BE-based) algorithms at different SNRs using $\mathrm{MSE}(\{\alpha_{\ell}\}_{\ell=1}^{L})$, $\mathrm{MSE}(\{\theta_{\mathrm{RX}, \ell}\}_{\ell=1}^{L})$, $\mathrm{MSE}(\{\theta_{\mathrm{TX}, \ell}\}_{\ell=1}^{L})$, and $\mathrm{NMSE}(\mathbf{H})$. The BE hard thresholding technique was proposed in \cite{doi:10.1137/110838509}, which was applied to the orthogonal matching pursuit (OMP) algorithm \cite{4385788}. In this paper, we apply the BE hard thresholding technique to the GraSP algorithm, which results in the BEGraSP algorithm. In this example, $B_{\mathrm{RX}}=B_{\mathrm{TX}}=256$ for the BMS-based and BE-based algorithms. We assume that $L=8$ and $\mathbf{H}$ is deterministic where $\alpha_{\ell}=(0.8+0.1(\ell-1))e^{j\frac{\pi}{4}(\ell-1)}$. However, $\{\theta_{\mathrm{RX}, \ell}\}_{\ell=1}^{L}$ and $\{\theta_{\mathrm{TX}, \ell}\}_{\ell=1}^{L}$ vary from simulation to simulation, which are either widely spread (Fig. \ref{figure_3a_alt}) or closely spread (Fig. \ref{figure_3b_alt}). In Figs. \ref{figure_3a_alt} and \ref{figure_3b_alt}, the notion of widely and closely spread paths refer to the fact that the minimum $2$-norm distance between the paths are either relatively far or close, i.e., $\min_{i\neq j}\|(\theta_{\mathrm{RX}, i}-\theta_{\mathrm{RX}, j}, \theta_{\mathrm{TX}, i}-\theta_{\mathrm{TX}, j})\|_{2}$ of Fig. \ref{figure_3a_alt}, which is $\sqrt{(\pi/18)^{2}+(\pi/18)^{2}}$, is greater than that of Fig. \ref{figure_3b_alt}, which is $\sqrt{(\pi/36)^{2}+(\pi/36)^{2}}$. The path gains, AoAs, and AoDs are assumed to be deterministic because the CRB is defined for deterministic parameters only \cite{poor2013introduction}. A variant of the CRB for random parameters is known as the Bayesian CRB, but adding the Bayesian CRB to our work is left as a future work because applying the Bayesian CRB to nonlinear measurements, e.g., one-bit ADCs, is not as straightforward.

According to Figs. \ref{figure_3a_alt} and \ref{figure_3b_alt}, the BMS-based algorithms succeed to estimate both widely spread and closely spread paths, whereas the BE-based algorithms fail to estimate closely spread paths. This follows because the BE hard thresholding technique was derived based on the assumption that $\mathrm{supp}(\mathbf{x}^{*})$ is widely spread. In contrast, the BMS hard thresholding technique is proposed based on Conjecture \ref{conjecture} without any assumption on $\mathrm{supp}(\mathbf{x}^{*})$. This means that when $\mathrm{supp}(\mathbf{x}^{*})$ is closely spread, the BE hard thresholding technique cannot properly estimate $\mathrm{supp}(\mathbf{x}^{*})$ because the BE hard thresholding technique, by its nature, excludes the elements near the maximum element of $\mathbf{x}^{*}$ from its potential candidate. The BMS hard thresholding technique, in contrast, uses the elements near the maximum element of $\mathbf{x}^{*}$ to construct the by-product testing set only, i.e., Line 4 of Algorithm \ref{bms}. Therefore, the BMS-based algorithms are superior to the BE-based algorithms when the paths are closely spread. The Cram\'er-Rao bounds (CRBs) of $\mathrm{MSE}(\{\alpha_{\ell}\}_{\ell=1}^{L})$, $\mathrm{MSE}(\{\theta_{\mathrm{RX}, \ell}\}_{\ell=1}^{L})$, and $\mathrm{MSE}(\{\theta_{\mathrm{TX}, \ell}\}_{\ell=1}^{L})$ are provided, which were derived in \cite{8335511}. The gaps between the MSEs and their corresponding CRBs can be interpreted as a performance limit incurred by the discretized AoAs and AoDs. To overcome such limit, the AoAs and AoDs must be estimated based on the off-grid method, which is beyond the scope of this paper.

In addition, note that $\mathrm{MSE}(\{\alpha_{\ell}\}_{\ell=1}^{L})$ and $\mathrm{NMSE}(\mathbf{H})$ worsen as the SNR enters the high SNR regime. To illustrate why $\mathbf{x}^{*}$ cannot be estimated in the high SNR regime in one-bit ADCs, note that
\begin{align}
\mathrm{Q}(\sqrt{\rho}\mathbf{A}\mathbf{x}^{*}+\mathbf{n})&\approx\mathrm{Q}(\sqrt{\rho}\mathbf{A}\mathbf{x}^{*})\notag\\
                                                          &=\mathrm{Q}(\sqrt{\rho}\mathbf{A}c\mathbf{x}^{*})
\end{align}
in the high SNR regime with $c>0$, which means that $\mathbf{x}^{*}$ and $c\mathbf{x}^{*}$ are indistinguishable because the magnitude information is lost by one-bit ADCs. The degradation of the recovery accuracy in the high SNR regime with one-bit ADCs is an inevitable phenomenon, as observed from other previous works on low-resolution ADCs \cite{8171203, 7931630, 8310593, 8320852, 8683652}.

In Figs. \ref{figure_4} and \ref{figure_5}, we compare the performance of Algorithms \ref{bmsgrasp}, \ref{bmsgrahtp}, and other estimators when $\mathbf{H}$ is random. The Bernoulli Gaussian-GAMP (BG-GAMP) algorithm \cite{8171203} is an iterative approximate MMSE estimator of $\mathbf{x}^{*}$, which was derived based on the assumption that $x_{i}^{*}$ is distributed as $\mathcal{CN}(0, 1)$ with probability $L/B$ but zero otherwise, namely the BG distribution. The fast iterative shrinkage-thresholding algorithm (FISTA) \cite{doi:10.1137/080716542} is an iterative MAP estimator of $\mathbf{x}^{*}$, which was derived based on the assumption that the logarithm of the PDF of $\mathbf{x}^{*}$ is $g_{\mathrm{FISTA}}(\mathbf{x})=-\gamma\|\mathbf{x}\|_{1}$ ignoring the constant factor, namely the Laplace distribution. Therefore, the estimate of $\mathbf{x}^{*}$ is
\begin{equation}\label{fista}
\underset{\mathbf{x}\in\mathbb{C}^{B}}{\mathrm{argmax}}\ (f(\mathbf{x})+g_{\mathrm{FISTA}}(\mathbf{x})),
\end{equation}
which is solved using the accelerated proximal gradient descent method \cite{doi:10.1137/080716542}. The regularization parameter $\gamma$ is selected so that the expected sparsity of \eqref{fista} is $3L$ for a fair comparison, which was suggested in \cite{JMLR:v14:bahmani13a}. In this example, $L=4$, whereas $B_{\mathrm{RX}}$ and $B_{\mathrm{TX}}$ vary from algorithm to algorithm. In particular, we select $B_{\mathrm{RX}}=B_{\mathrm{TX}}=256$ for Algorithms \ref{bmsgrasp}, \ref{bmsgrahtp}, and the FISTA, whereas $B_{\mathrm{RX}}=M$ and $B_{\mathrm{TX}}=N$ for the BG-GAMP algorithm.

In Fig. \ref{figure_4}, we compare the accuracy of Algorithms \ref{bmsgrasp}, \ref{bmsgrahtp}, and other estimators at different SNRs using $\mathrm{NMSE}(\mathbf{H})$. According to Fig. \ref{figure_4}, Algorithms \ref{bmsgrasp} and \ref{bmsgrahtp} outperform the BG-GAMP algorithm and FISTA as the SNR enters the medium SNR regime. The accuracy of the BG-GAMP algorithm is disappointing because the mismatch in \eqref{virtual_channel_representation} is inversely proportional to $B_{\mathrm{RX}}$ and $B_{\mathrm{TX}}$. However, increasing $B_{\mathrm{RX}}$ and $B_{\mathrm{TX}}$ is forbidden because the BG-GAMP algorithm diverges when $\mathbf{A}$ is highly coherent. The accuracy of the FISTA is disappointing because the Laplace distribution does not match the distribution of $\mathbf{x}^{*}$. Note that \eqref{map}, which is the basis of Algorithms \ref{bmsgrasp} and \ref{bmsgrahtp}, is indeed the MAP estimate of $\mathbf{x}^{*}$, which is in contrast to the FISTA. According to Fig. \ref{figure_4}, $\mathrm{NMSE}(\mathbf{H})$ worsens as the SNR enters the high SNR regime, which follows from the same reason as in Figs. \ref{figure_3a_alt} and \ref{figure_3b_alt}.


In Fig. \ref{figure_5}, we compare the achievable rate lower bound of Algorithms \ref{bmsgrasp}, \ref{bmsgrahtp}, and other estimators at different SNRs when the precoders and combiners are selected based on $\tilde{\mathbf{H}}$. The achievable rate lower bound shown in Fig. \ref{figure_5} is presented in \cite{8171203}, which was derived based on the Bussgang decomposition \cite{bussgang1952crosscorrelation} in conjunction with the fact that the worst-case noise is Gaussian. According to Fig. \ref{figure_5}, Algorithms \ref{bmsgrasp} and \ref{bmsgrahtp} outperform the BG-GAMP algorithm and FISTA, which is consistent with the result in Fig. \ref{figure_4}.

\begin{figure}[t]
\centering
\includegraphics[width=1\columnwidth]{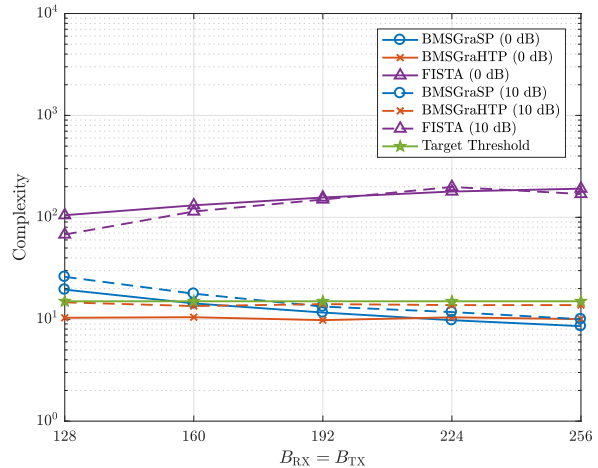}
\caption{Normalized complexity vs. $B_{\mathrm{RX}}=B_{\mathrm{TX}}$ where $M=N=64$, $T=80$, and $L=4$ at $\mathrm{SNR}=0$ dB and $\mathrm{SNR}=10$ dB.}\label{figure_6}
\end{figure}

In Fig. \ref{figure_6}, we compare the complexity of Algorithms \ref{bmsgrasp}, \ref{bmsgrahtp}, and other estimators at different $B_{\mathrm{RX}}$ and $B_{\mathrm{TX}}$ when $\mathbf{H}$ is random. To analyze the complexity, note that Algorithms \ref{bmsgrasp}, \ref{bmsgrahtp}, and the FISTA require $h(\mathbf{x})$ and $\nabla h(\mathbf{x})$ at each iteration, whose bottlenecks are $\mathbf{A}\mathbf{x}$ and $\mathbf{A}^{\mathrm{H}}\mathbf{c}$, respectively, while the BG-GAMP algorithm requires $\mathbf{A}\mathbf{x}$ and $\mathbf{A}^{\mathrm{H}}\mathbf{c}$ at each iteration. Therefore, the complexity is measured based on the number of complex multiplications performed to compute $\mathbf{A}\mathbf{x}$ and $\mathbf{A}^{\mathrm{H}}\mathbf{c}$, which are implemented based on the FFT. In this example, $L=4$, whereas SNR is either $0$ dB or $10$ dB.

In this paper, the complexity of the BG-GAMP algorithm is used as a baseline because the BG-GAMP algorithm is widely used. The normalized complexity is defined as the number of complex multiplications performed divided by the per-iteration complexity of the BG-GAMP. For example, the normalized complexity of the FISTA with $B_{\mathrm{RX}}=B_{\mathrm{TX}}=256$ is $160$ when the complexity of the FISTA with $B_{\mathrm{RX}}=B_{\mathrm{TX}}=256$ is equivalent to the complexity of the $160$-iteration BG-GAMP algorithm with $B_{\mathrm{RX}}=B_{\mathrm{TX}}=256$. In practice, the BG-GAMP algorithm converges in $15$ iterations when $\mathbf{A}$ is incoherent \cite{6556987}. In this paper, an algorithm is said to be as efficient as the BG-GAMP algorithm when the normalized complexity is below the target threshold, which is $15$. As a sidenote, our algorithms, namely the BMSGraSP and BMSGraHTP, requires $2.1710$ and $2.0043$ iterations in average, respectively, across the entire SNR range.

According to Fig. \ref{figure_6}, the complexity of the FISTA is impractical because the objective function of \eqref{fista} is a high-dimensional function defined on $\mathbb{C}^{B}$ where $B\gg MN$. In contrast, the complexity of Algorithms \ref{bmsgrasp} and \ref{bmsgrahtp} is dominated by \eqref{line_5}, whose objective function is a low-dimensional function defined on $\mathbb{C}^{|\mathcal{I}|}$ where $|\mathcal{I}|=O(L)$. The normalized complexity of Algorithms \ref{bmsgrasp} and \ref{bmsgrahtp} is below the target threshold when $B_{\mathrm{RX}}\geq 192$ and $B_{\mathrm{TX}}\geq 192$. Therefore, we conclude that Algorithms \ref{bmsgrasp} and \ref{bmsgrahtp} are as efficient as the BG-GAMP algorithm when $B_{\mathrm{RX}}\gg M$ and $B_{\mathrm{TX}}\gg N$.

\section{Conclusions}\label{conclusion}
In the mmWave band, the channel estimation problem is converted to a sparsity-constrained optimization problem, which is NP-hard to solve. To approximately solve sparsity-constrained optimization problems, the GraSP and GraHTP algorithms were proposed in CS, which pursue the gradient of the objective function. The GraSP and GraHTP algorithms, however, break down when the objective function is ill-conditioned, which is incurred by the highly coherent sensing matrix. To remedy such break down, we proposed the BMS hard thresholding technique, which is applied to the GraSP and GraHTP algorithms, namely the BMSGraSP and BMSGraHTP algorithms, respectively. Instead of directly hard thresholding the gradient of the objective function, the BMS-based algorithms test whether an index is the ground truth index or the by-product of another index. We also proposed an FFT-based fast implementation of the BMS-based algorithms, whose complexity is reduced from $O(M^{4})$ to $O(M^{2}\log M)$. In the simulation, we compared the performance of the BMS-based, BE-based, BG-GAMP, and FISTA algorithms from different aspects in terms of the accuracy, achievable rate, and complexity. The BMS-based algorithms were shown to outperform other estimators, which proved to be both accurate and efficient. Our algorithms, however, addressed only the flat fading scenario, so an interesting future work would be to propose a low-complexity channel estimator capable of dealing with the wideband scenario.

\bibliographystyle{IEEEtran}
\bibliography{refs_all}

\end{document}